\documentclass[a4paper,12pt]{article}
\usepackage{amssymb,amsfonts,amsmath,mathtext,cite,enumerate,float,color}
\usepackage{amssymb,amsthm,latexsym, braket, euscript,amscd, authblk}

\usepackage[russian,english]{babel}
\usepackage[utf8]{inputenc}

\title{On quantum tomography on locally compact groups}

\author[1]{G.G.Amosov}

\affil[1]{Steklov Mathematical Institute of Russian Academy of Sciences}

\begin{document}

\maketitle

\begin{abstract}
We introduce quantum tomography on locally compact Abelian groups $G$. A linear map from the set of quantum states on the $C^*$-algebra $A(G)$ generated by the projective unitary representation of $G$ to the space of characteristic functions is constructed. The dual map determining symbols of quantum observables from $A(G)$ is derived. Given a characteristic function of a state the quantum tomogram consisting a set of probability distributions is introduced. We provide three examples in which $G={\mathbb R}$ (the optical tomography), $G={\mathbb Z}_n$ (corresponding to measurements in mutually unbiased bases) and $G={\mathbb T}$ (the tomography of the phase). As an application we have calculated the quantum tomogram for the output states of quantum Weyl channels.
\end{abstract}

\section{Introduction}

The problem of measuring obseravables associated with linear combinations of the position and momentum operators was experimentally solved by means of homodyne tomography introduced in the pioneering paper \cite{Raymer}. A little later, an approach was proposed to calculate the optical tomogram without an intermediate stage involving the Wigner function \cite{DAriano}. In \cite{Manko, Manko2}, an ambitious program was set for the formulation of quantum mechanics in terms of quantum tomography. In this formulation, the role of quantum states is played by tomograms. In this case, the quantum observables act as generalized functions on the space of test functions consisting of tomograms \cite{Amo1, Amo2, Amo3}.

An optical quantum tomogram is a set of probability distributions on $\mathbb R$. It means that the measurement results belong to $\mathbb R$. A real line $\mathbb R$ is an Abelian group with respect to addition operation. It is naturally to extend quantum tomography to arbitrary locally compact groups. The motivation for setting such a problem is that some of the information is lost during the experimental measurement, so that the corresponding probability distributions are determined rather not on $\mathbb R$, but on some discrete group. Also, the measurements themselves are made not by a continuum, but by a finite number. The problem of the accuracy of the reconstruction of a quantum state from the partial information about its tomogram is extremely relevant \cite{Guta}. Earlier the Wigner function was introduced on Lie groups \cite {Ali} but no tomography (sets of probability distributions determining a state) was defined. In our approach, we immediately limit the allowable amount of information by moving from $\mathbb R$ to some specific group. Our requirements for the corresponding group are limited by local compactness.

This paper is organized as follows. At first, we introduce a projective unitary representation of the group $G$ in the Hilbert space consisting of functions on $G$. Then, we define characteristic functions of states on the $C^*$-algebra $A(G)$ generated by the representation. Here the Parseval identity is proved. On the second step, we determine symbols of observables from $A(G)$. Finally, the quantum tomogram of a state (the set of probability distributions) is introduced. Then, three examples including $G={\mathbb R}$, $G={\mathbb Z}_n$ and $G={\mathbb T}$ are presented. The last part is devoted to applications to the study of the channels being a convex sum of actions fulfilled  by Weyl operators. The quantum tomograms of the output states of Weyl channels are derived in the evident form.

\section{Tomography on groups}

Let $G$ be a locally compact Abelian group with the Haar measure $\mu $. Denote $\hat G$ its dual group with the Haar measure $\nu $. Let us define a projective unitary representation $\pi$ of $\hat G\times G$ in the Hilbert space ${\mathcal H}=L^2(G)=\{f:\ \int \limits _G|f(g)|^2d\mu (g)<+\infty \}$ by the formula
$$
(\pi (\chi , g)f)(a)=\chi (a)f(a+g).
$$
It is straightforward to check that
\begin{equation}\label{tozhd}
\pi (\chi '\chi  , g'+g)=\chi (g')^*\pi (\chi ', g')\pi (\chi ,g),
\end{equation}
$\chi ,\chi '\in \hat G,\ g,g'\in G$. 

In what follows we put $(e^{i\varphi })^{1/2}\equiv e^{i\varphi /2}$ for $\varphi \in [0,2\pi )$.
Consider the Hilbert space ${\mathcal K}=L^2(\nu \times \mu)$ consisting of functions on $\hat G\times G$.

{\bf Proposition 1.} {\it Suppose that $\{f_j\}$ is an orthonormal basis in $\mathcal H$. Then, the functions
\begin{equation}\label{basis}
F_{jk}(\chi ,g)=\braket {f_j,\pi (\chi ,g)f_k},\ \chi \in \hat G,\ g\in G,
\end{equation}
form the orthonormal basis in $\mathcal K$.
}

Proof.

Given $F\in L^1(\hat G\times G)$ put 
$$
\hat F(\chi ,g)=\int \limits _{\hat G\times G}\chi (g')\chi '(g)^*F(\chi ',g')d\nu (\chi ')d\mu (g'),\ \chi \in \hat G,\ g\in G,
$$
with the inverse transform given by
$$
F(\chi ,g)=\int \limits _{\hat G\times G}\chi (g')^*\chi '(g)\hat F(\chi ',g')d\nu (\chi ')d\mu (g').
$$
It is known that the formulas above can be extended to $F\in L^2(\hat G\times G)$ due to the Parseval identity \cite{Reiter}.

Taking into account the inclusion $F_{jk}\in L^2(\hat G\times G)$ because $f_j,f_k\in L^2(\hat G\times G)$ we obtain 
$$
F_{jk}(\chi ,g)=\braket {f_j,\pi (\chi ,g)f_k}=\int \limits _Gf_j^*(g')\chi (g')f_k(g'+g)d\mu (g')=
$$
$$
\int \limits _{\hat G\times G}f_j^*(h)\chi (g')\chi '(g)^*\chi '(g')\hat f_k(\chi ')d\nu (\chi ')d\mu (g')=
$$
$$
\widehat {[\chi '(g')\hat {f_k}(\chi ')f_j^*(g')]}(\chi ,g).
$$
Since the functions
$$
f_{jk}(\chi ,g)=\chi (g)\hat f_k(\chi )f_j^*(g)
$$
form the orthonormal basis in $\mathcal K$
we get that
$$
F_{jk}=\hat f_{jk}
$$
is also an orthonormal basis in $\mathcal K$.

$\Box$

Denote $\mathfrak {S}_2({\mathcal H})$ the space of Hilbert-Schmidt operators in $\mathcal H$. The convex set of quantum states (positive unit trace operators) ${\mathfrak S}({\mathcal H})\subset \mathfrak {S}_2({\mathcal H})$.
Let us define a map $\Phi $ on rank one operators by the formula
\begin{equation}\label{map}
[\Phi (\ket {f}\bra {h})](\chi ,g)=\braket {h,\pi (\chi ,g)f},\ f,h\in {\mathcal H},\ g\in G,\ \chi \in \hat G.
\end {equation}

{\bf Theorem (the Parseval identity).} {\it Formula (\ref {map}) determines a linear map $\Phi :\mathfrak {S}_2({\mathcal H})\to {\mathcal K}$ such that
$$
||\Phi (\rho )||^2=Tr(\rho ^*\rho ),\ \rho \in \mathfrak {S}_2({\mathcal H}).
$$
}

Proof.

Pick up the orthogonal basis $\{f_j\}$ in $\mathcal H$. Then, any $\rho \in \mathfrak {S}_2({\mathcal H})$ can be represented in the form
\begin{equation}\label{ryad}
\rho =\sum \limits _{j,k}\lambda _{jk}\ket {f_k}\bra {f_j}
\end{equation}
and the series converges in the Hilbert-Schmidt norm. Applying $\Phi $ to (\ref {ryad}) we obtain
$$
[\Phi (\rho )](\chi ,g)=\sum \limits _{j,k}\lambda _{jk}F_{jk}(\chi ,g),
$$
where $F_{jk}$ is the orthonormal basis in $\mathcal K$ due to Proposition 1.

$\Box $

{\bf Definition 1.} {\it Given a unit vector $f \in L^2({\mathcal H})$ the function $\Phi (\ket {f}\bra {f})$ is
said to be a characteristic function of the pure state $\ket {f}\bra {f}$.}

Consider the $C^*$-algebra $A(G)$ generated by all operators $\pi (\chi ,g),\ \chi \in \hat G,\ g\in G$, equipped with the operator norm in $\mathcal H$.

{\bf Definition 2.} {\it The functional $W_x\in C(\hat G\times G)^*$ is said to a symbol of the operator $x\in A(G)$ if
its action results in the mean value of $x$ in the sense
$$
<W_x, \Phi (\ket {f}\bra {f})>=\braket {f,xf},
$$
for all $f\in \mathcal H$.}

{\bf Proposition 2.} {\it The functional $W_{\chi _0,g_0}(\chi ,g)=\delta (\chi -\chi _0,g-g_0)\in C(\hat G\times G)^*$ acting by the formula
$$
<W_{\chi _0,g_0},F>=F(\chi _0,g_0),\ F\in C(\hat G\times G),
$$
is a symbol of operator $\pi (\chi _0,g_0)$.}

Proof.

It immediately follows from (\ref {map}).

$\Box $

Fix $\chi \in \hat G$ and $g\in G$ and consider the set $G_{\chi ,g}=\{(\chi ',g'):\ \chi '(g)=\chi (g')\}$.

{\bf Lemma 1.} {\it $G_{\chi ,g}$ is a subgroup of $\hat G\times G$.}

Proof.

If $\chi '(g)=\chi (g')$ and $\chi ''(g)=\chi (g'')$, then 
$$
(\chi '\chi '')(g)=\chi '(g)\chi ''(g)=\chi (g')\chi (g'')=\chi (g'+g'').
$$

$\Box $

{\bf Lemma 2.} {\it The map $(\chi ',g')\in G_{\chi ,g}\to [\chi '(g')]^{1/2}\pi (\chi ',g')$ is a unitary representation
of $G_{\chi ,g}$ in ${\mathcal H}$.}

Proof.

Given $f\in \mathcal H$ we get
\begin{equation}\label{vspom}
([\chi '(g')\chi (g)]^{1/2}\pi (\chi ',g')\pi (\chi ,g)f)(a)=[\chi '(g')\chi (g)]^{1/2}\chi '(a)\chi (a+g')f(a+g+g').
\end{equation}
Taking into account
$$
\chi '(g)=\chi (g')
$$
due to Definition of $G_{\chi ,g}$ results in
\begin{equation}\label{vspom2}
[\chi '(g')\chi (g)]^{1/2}\chi (g')=[\chi '(g')\chi (g)]^{1/2}[\chi '(g)]^{1/2}\chi ^{1/2}(g')=[\chi '\chi (g+g')]^{1/2}.
\end{equation}
Substituting (\ref {vspom2}) to (\ref {vspom}) we obtain
$$
([\chi '(g')\chi (g)]^{1/2}\pi (\chi ',g')\pi (\chi ,g)f)(a)=([\chi '\chi (g+g')]^{1/2}\pi (\chi '\chi ,g'+g)f)(a).
$$
The result follows.

$\Box $

Given a unit vector $f\in {\mathcal H}$ let us consider the restriction $\Phi (\ket {f}\bra {f})$ defined by (\ref {map}) on $G_{\chi ,g}$ given by the formula
\begin{equation}\label{charact}
F_f(\chi  ',g')=\chi '(g')^{1/2}\braket {f,\pi (\chi ',g')f},\ (\chi ',g')\in G_{\chi ,g}.
\end{equation}

{\bf Lemma 3.} {$F_f$ is a positive definite function on $G_{\chi ,g}$ for any fixed $\chi \in \hat G,g\in G$.}

Proof.

Take  $(\chi _k,g_k)\in G_{\chi ,g}$ and $\lambda _k\in {\mathbb C}$, then
$$
\sum \limits _{j,k}\lambda _j^*\lambda _k[(\chi _j\chi _k^*)(g_j-g_k)]^{1/2}\braket {f,\pi (\chi _j\chi _k^*,g_j-g_k)f}=
$$
$$
\braket {\sum \limits _j\lambda _j[\chi _j(g_j)]^{1/2}\pi (\chi _j,g_j)f,
\sum \limits _k\lambda _k[\chi _k(g_k)]^{1/2}\pi (\chi _k,g_k)f}
$$
because 
$$
[(\chi _j\chi _k^*)(g_j-g_k)]^{1/2}\pi (\chi _j\chi _k^*,g_j-g_k)=\chi _j(g_j)\chi _k(g_k)\pi (\chi _k^*,-g_k)\pi (\chi _j,g_j)
$$
in virtue of (\ref {tozhd}) and
$$
\chi _k(g_j)=\chi _j(g_k)
$$
due to Definition of $G_{\chi ,g}$.

$\Box $

Notice that the group $G_{\chi ,g}$ is self-dual such that $\hat G_{\chi ,g}\approx G_{\chi ,g}$. Denote
$(\chi ',g')\to X_{\chi ',g'}\in \hat G_{\chi ,g}$ the corresponding isomorphism.

{\bf Proposition 3.} {\it Given a unit vector $f\in \mathcal H$ and fixed $\chi \in \hat G,$ $g\in G$ there exists a probability measure $\mu _{\chi ,g}^f$ on $G_{\chi ,g}$ such that
$$
F_f(\chi ',g')=\int \limits _{G_{\chi ,g}}X_{\chi ',g'}(\chi '',g'')d\mu _{\chi ,g}^f(\chi '',g''),\ (\chi ',g')\in G_{\chi ,g}.
$$
}

Proof.

Formula (\ref {map}) implies that $\Phi (\ket {f}\bra {f})\in C(\hat G\times G)$ and $\Phi (\ket {f}\bra {f})(1,0)=1$.
Since the restriction of $F_f$ is positive definite due to Lemma 3 the result follows from the Bochner theorem.

$\Box $

{\bf Definition 3.} {\it The set of probability distributions
$\{\mu _{\chi ,g}^f,\ \chi \in \hat G,\ g\in G\}$ is said to be {\it a quantum tomogram} of a pure state $f$.}

{\bf Example 1.} $G={\mathbb R}$.

In the case, $\hat G$ is isomorphic to $G$ by means of the map $x\in G\to \chi _x\in \hat G$, where $\chi _x(y)=exp(ixy)$. Formula (\ref {charact}) reads
$$
F_f(x,y)=exp\left (\frac {ixy}{2}\right )\int \limits _{\mathbb R}exp(ixt)f(t+y)f^*(t)dt=
$$
$$
\int \limits _{\mathbb R}exp(ixt)f\left (t+\frac {y}{2}\right )f^*\left (t-\frac {y}{2}\right )dt
$$
and it is the characteristic function of a pure state $f$ \cite{Holevo}. 

The subgroup $G_{x,y}=\{x',y':\ xy'=x'y\}$ can be parametrized by $\varphi \in [0,2\pi )$ such that $x\sin \varphi =y\cos \varphi $ and
$$
G_{x,y}\equiv G_{\varphi }=\{x',y':\ x'=t\cos\varphi ,\ y'=t\sin \varphi ,\ t\in {\mathbb R}\}.
$$
In turn, the quantum tomogram of Definition 3 becomes
$$
\mu _{\varphi }^f(B)=\int \limits _B\omega _f(X,\varphi )dX,
$$
for all measurable $B\subset \mathbb R$, where
$$
\omega _f(X,\varphi )=\frac {1}{2\pi }\int \limits _{\mathbb R}exp(-iXt)F_f(t\cos \varphi ,t\sin\varphi )dt,
$$
$X\in {\mathbb R}$, is the optical tomogram of a pure quantum state $f$.

Notice that the unitary representation of Lemma 2 has the form
$$
G_{\varphi }\ni t\to exp(it(\cos\varphi q+\sin \varphi p)),
$$
where $q$ and $p$ are the position and momentum operators. Denote $f_{\varphi }(X)$ a wave function of the state $\ket {f}\bra {f}$ corresponding to the observable $\cos\varphi q+\sin\varphi p$. Then,
$$
\omega _f(X,\varphi )=|f_{\varphi }(X)|^2.
$$

{\bf Example 2.} {\it $G={\mathbb Z}_n$.} 

For the sake of simplicity we claim $n=p^N$ for some prime number $p$ what allows you to determine the division.

Analogously to Example 1 $\hat G$ is isomorphic to $G$ by means of the map $k\in {\mathbb Z}_n\to \chi _k\in \hat G$, where
$\chi _k(m)=exp\left (i\frac {2\pi km}{n}\right )$. Here the subgroups $G_{k,m}=\{k',m':\ km'=k'm\ mod\ n\}$. 

{\bf Lemma 4.} {\it The pair $(k,m)$ determining the subgroup $G_{k,m}$ 
can be parametrized by $l:\ 0\le l\le n$ such that 
$$
G_l=\{(kl,k),\ k\in {\mathbb Z}_n\},\ 0\le l\le n-1,\ l=\frac {k}{m},\ m\neq 0,
$$
and
$$
G_n=\{(k,0),\ k\in {\mathbb Z}_n\},\ m=0.
$$
}

Proof.

All elements of $G_l$ satisfy the condition of the corresponding $G_{k,m}$ such that $G_l\subset G_{k,m}$. Since $n=p^N$ we hat a devision operation and $G_l\equiv G_{k,m}$.

$\Box $

Denote $e_j(k)=\delta _{jk},\ j,k\in {\mathbb Z}_n$, the natural orthonormal basis in $\mathcal {H}=L^2({\mathbb Z}_n)$ and define two unitary operators $U$ in $V$ in $\mathcal H$ by the formula
$$
Ue_j=e^{i\frac {2\pi j}{n}}e_j,\ Ve_j=e_{j+1},\ j\in {\mathbb Z}_n.
$$
Following to \cite {Amo4} let us define unitary representations of $G_l,\ 0\le l\le n,$ as follows
$$
\pi _l(kl,k)=(U^lV)^k,\ k\in {\mathbb Z}_n,\ 0\le l\le n-1,
$$
\begin{equation}\label{preds}
\pi _n(k,0)=U^k,\ k\in {\mathbb Z}_n.
\end{equation}
Now the characteristic function of a pure state $f\in {\mathcal H}$ is given by
$$
F_f(lk,k)=\braket {f,(U^lV)^kf},\ F_f(k,0)=\braket {f,U^kf},
$$
$k\in {\mathbb Z}_n$,
and densities of the quantum tomogram equal
$$
\omega _f(j,l)=\frac {1}{n}\sum \limits _{k\in {\mathbb Z}_n}e^{-i\frac {2\pi jk}{n}}F_f(lk,k),\ 0\le l\le n-1,
$$
$$
\omega _f(j,n)=\frac {1}{n}\sum \limits _{k\in {\mathbb Z}_n}e^{-i\frac {2\pi jk}{n}}F_f(k,0),
$$
$j\in {\mathbb Z}_n$.

The eigenvectors $e_j^l$ of the operators $U^lV,\ 0\le l\le n-1,$ and $U,\ l=n,$ determined by the relation
$$
\ket {e_j^l}\bra {e_j^l}=\frac {1}{n}\sum \limits _{k=0}^{n-1}e^{i\frac {2\pi jk}{n}}(U^lV)^k,\ 0\le l\le n-1,
$$ 
$$
\ket {e_j^n}\bra {e_j^n}=\frac {1}{n}\sum \limits _{k=0}^{n-1}e^{i\frac {2\pi jk}{n}}U^k\equiv \ket {e_j}\bra {e_j},
$$
$j\in {\mathbb Z}_n$, form the full set of mutually unbiased bases satisfying
$$
|\braket {e_j^l,e_k^m}|^2=\frac {1}{n},\ l\neq m,\ j,k\in {\mathbb Z}_n.
$$
Using these bases the densities can be represented as
\begin{equation}\label{TOM}
\omega _f(j,l)=|\braket {e_j^l,f}|^2,\ j\in {\mathbb Z}_n,\ 0\le l\le n.
\end{equation}

{\bf Example 3.} $G={\mathbb T}=[0,2\pi)$ the circle group with the multiplication $+/2\pi $.

Using this group involves the measurement of the phase. Here $\hat G={\mathbb Z}$ and the duality map is given by the formula
$$
<n,\varphi >=exp(in\varphi ),\ \varphi \in G,\ n\in \hat G.
$$
The subgroups $G_{n,\varphi }=\{(n',\varphi'):\ \varphi n'=\varphi 'n\ mod\ 2\pi\}$.

{\bf Lemma 5.} {\it The subgroups $G_{n,\varphi }$ can be parametrized as follows
$$
G_{n,\varphi }=\{(n',\varphi '):\ \varphi '=n'\theta\ mod\ 2\pi ,\ \theta =\frac {\varphi}{n} \ mod\ 2\pi\}\equiv G_{\theta },\ n\neq 0,\ \theta \in {\mathbb T},
$$
$$
G_{0,\varphi }=\{(0,\varphi),\ \varphi \in {\mathbb T}\}\equiv G_*.
$$
}

Proof.

It immediately follows from the definition.

$\Box $

Taking into account Lemma 2 let us define a unitary representation of each $G_{\theta },\ \theta \in \mathbb {T},$ in the Hilbert space ${\mathcal H}=L^2({\mathbb T})$ as follows
$$
(\pi _{\theta }(n,n\theta )f)(\psi )=e^{in \left (\psi +\frac {n\theta}{2}\right )}f(\psi +n\theta ),\ (n,n\theta )\in G_{\theta },\ 
 \theta \in \mathbb {T},
$$
$$
(\pi _*(0,\varphi )f)(\psi )=f(\psi +\varphi ),\ (0,\varphi )\in G_*,\ f\in {\mathcal H}.
$$
For the characteristic function $F_f$ it results in
$$
F_f(n,n\theta )=\braket {f,\pi _{\theta }(n,n\theta )f}=\int \limits _0^{2\pi}e^{in\psi }f\left (\psi +\frac {n\theta }{2}\right )f^*\left (\psi -\frac {n\theta }{2}\right )d\psi ,
$$
$(n,n\theta )\in G_{\theta }$,
$$
F_f(0,\varphi )=\int \limits _0^{2\pi }f(\psi +\varphi )f^*(\psi)d\psi ,\ (0,\varphi )\in G_*.
$$
Now the quantum tomogram is given by the Fourier series
$$
\omega _f(n,\theta )=\frac {1}{2\pi }\sum \limits _{m\in {\mathbb Z}}e^{-inm}F_f(m,m\theta ),\ \theta \in {\mathbb T},
$$
together with the Fourier transform
$$
\omega _f(n,*)=\frac {1}{2\pi }\int \limits _{0}^{2\pi }e^{-in\varphi }F_f(0,\varphi )d\varphi ,
$$
$n\in {\mathbb Z}$. 

\section{Application to the study of mixed unitary quantum channels}

Put $G={\mathbb Z}_n$ and $n=p^N$ for some prime number $p$ as in the previous section. Given a probability distribution $\{q_{k,m},\ k,m\in {\mathbb Z}_n\}$ let us define a mixed unitary quantum channel $\Phi :\mathfrak {S}(\mathcal {H})\to \mathfrak {S}(\mathcal {H})$ by the formula
\begin{equation}\label{chan}
\Phi (\rho )=q_{0,0}\rho+\sum \limits _{l=0}^{n-1}\sum \limits _{k=1}^{n-1}q_{kl,k}\pi _l(kl,k)\rho \pi _l(kl,k)^*+
\sum \limits _{k=1}^{n-1}q_{k,0}\pi _n(k,0)\rho\pi _n(k,0)^*,
\end{equation}
$\rho \in {\mathfrak S}(\mathcal {H}),$
where unitary operators $\pi _l(g)$ are determined by (\ref {preds}). 
It follows from Lemma 4 that
$$
{\mathbb Z}_n\times {\mathbb Z}_n=\oplus _{l=0}^nG_l.
$$
Our goal is to fulfil quantum tomography of the output state $\rho _{out}=\Phi (\rho )$ for the case of a pure input state $\rho =\ket {f}\bra {f},\ f \in {\mathcal H},\ ||f||=1$.

Put
$$
Q_{ml}=\sum \limits _{k=0}^{n-1}q_{m+kl,k},
$$
\begin{equation}\label{marginal}
Q_{mn}=\sum \limits _{k=0}^{n-1}q_{m+k,m},\ 0\le m,l\le n-1.
\end{equation}
Notice that (\ref {marginal}) determine marginal probability distributions for $\{q_{k,m}\}$. Under some assumptions the entropy of one of the distributions (\ref {marginal}) gives the minimal output von Neumann entropy for the channel (\ref {chan}). In the case, it allows to calculate a classical capacity \cite{Amo5}.

{\bf Proposition 4.} {\it The characteristic function of the output state $\rho _{out}=\Phi (\ket {f}\bra{f})$ has the form
$$
F_{\rho _{out}}(kl,k)=\left (\sum \limits _{m=0}^{n-1}Q_{ml}e^{\frac {i2\pi km}{n}}\right )\braket {f,(U^lV)^kf},
$$
$$
F_{\rho _{out}}(k,0)=\left (\sum \limits _{m=0}^{n-1}Q_{mn}e^{\frac {i2\pi km}{n}}\right )\braket {f,U^kf},\ 0\le k\le n-1.
$$
}

Proof.

It follows from (\ref {preds}) that
$$
\pi _{l'}(ml',m)^*\pi _{l}(kl,k)\pi _{l'}(ml',m)=e^{\frac {2\pi ikm}{n}(l'-l)}\pi _{l}(kl,k),
$$ 
$$
\pi _{l'}(ml',m)^*\pi _n(k,0)\pi _{l'}(ml',m)=e^{\frac {2\pi ikm}{n}}\pi _n(k,0),\ 0\le l,l'\le n-1.
$$
Resolving the equation
$$
m(l'-l)=k
$$
we obtain
$$
ml'=k+ml
$$

$\Box $

Using the quantum tomogram of a state $\rho _{in}=\ket {f}\bra{f}$ given by (\ref {TOM}) we can find the tomogram of $\rho _{out}=\Phi (\rho _{in})$.

{\bf Corollary.} {\it The quantum tomogram of $\rho _{out}$ is determined by the convolution as follows
$$
\omega _{\rho _{out}}(j,l)=\sum \limits _{m=0}^{n-1}Q_{m,l}\omega _f(n-1-m,l),\ j\in {\mathbb Z}_n,\ 0\le l\le n.
$$
}

Proof.

It suffices to remember that a multiplication transfer to a convolution under the Fourier transform.

$\Box$

\section{Conclusion}

We introduced quantum tomography for states on the $C^*$-algebra generated by the projective unitary representation of a locally compact Abelian group $G$. The corresponding symbols of quantum observables allowing to calculate mean values are determined. The construction is provided by three Examples of $G={\mathbb R}$, ${\mathbb Z}_n$ and ${\mathbb T}$, which correspond to the cases of measurements of the homodyne quadrature, in mutually unbiased bases and the phases, respectively. We apply these techniques to study of Weyl channels. It is shown that the quantum tomogram of the output state of the Weyl channel can be represented as a convolution of the tomogram of the input state with the probability distribution determined by the channel.

\end{document}